\documentclass[]{aastex631}

\newcommand\degree{^o}
\def\las{\mathrel{\hbox{\rlap{\hbox{\lower3pt\hbox{$\sim$}}}\hbox{\raise2pt\hbox{$<$}}}}}
\def\gas{\mathrel{\hbox{\rlap{\hbox{\lower3pt\hbox{$\sim$}}}\hbox{\raise2pt\hbox{$>$}}}}}
\usepackage{xcolor}

\usepackage{amsmath}

\begin{document}

\title{A limit on the mass of the Taurid Resonant Swarm at sub-100~meter sizes}

\author[0000-0002-1914-5352]{Paul Wiegert}
\affiliation{Department of Physics and Astronomy, The University of Western Onta
rio, London, Ontario, Canada}
\affiliation{Institute for Earth and Space Exploration, The University of Wester
n Ontario, London, Ontario, Canada}

\author[0000-0003-4166-8704]{Denis Vida}
\affiliation{Department of Physics and Astronomy, The University of Western Onta
rio, London, Ontario, Canada}
\affiliation{Institute for Earth and Space Exploration, The University of Wester
n Ontario, London, Ontario, Canada}

\author[0000-0002-1203-764X]{David Clark}
\affiliation{Department of Physics and Astronomy, The University of Western Onta
rio, London, Ontario, Canada}
\affiliation{Institute for Earth and Space Exploration, The University of Wester
n Ontario, London, Ontario, Canada}

\author[0000-0002-9572-1200]{Auriane Egal}
\affiliation{IMCCE, Observatoire de Paris, PSL Research University, CNRS UMR 802
8, Sorbonne Universit\'e, Universit\'e de Lille, 7 av. Denfert-Rochereau, Paris,
 75014, France}
\affiliation{Plan\'etarium de Montr\'eal, Espace pour la Vie, 4801 av. Pierre-de
 Coubertin, Montréal, H1V 3V4, Québec, Canada}
\affiliation{Department of Physics and Astronomy, The University of Western Onta
rio, London, Ontario, Canada}
\affiliation{Institute for Earth and Space Exploration, The University of Wester
n Ontario, London, Ontario, Canada}

\author[0000-0002-1341-0952]{Richard Wainscoat}
\affiliation{Institute for Astronomy, University of Hawaii, 2680 Woodlawn Drive,
 Honolulu HI 96822, USA}

\author[0000-0002-0439-9341]{Robert Weryk}
\affiliation{Department of Physics and Astronomy, The University of Western Onta
rio, London, Ontario, Canada}

\correspondingauthor{Paul Wiegert}
\email{pwiegert@uwo.ca}



\begin{abstract}

We report on a pencil-beam survey of the Taurid Swarm, a possible concentration of bodies in the Taurid meteoroid stream associated with the 7:2 mean-motion resonance with Jupiter.  Canada-France-Hawaii Telescope MegaCam observations reaching apparent magnitudes of 24.5 in the \textit{gri} filter were taken over three nights. Rates of motion on the sky allowed for the quick elimination of main-belt objects from the over 1000 moving sources seen. Eight candidates with on-sky rates of motion consistent with Taurids were detected, but seven were subsequently shown to be non-Taurids (Hungarias, Mars-crossers, etc). One object might be a 60~m class Taurid but not enough data was collected and its orbit remains ambiguous. Our results are consistent with no Taurid Swarm members observed, and an upper limit of fewer than $3 \times 10^3- 3 \times 10^4$ objects down to $H=25.6\pm0.3$ (diameter of $47_{-13}^{+29}$~m assuming a 2P/Encke-like albedo) at the 95\% confidence level. While meteor observations have confirmed the Taurid Swarm's existence at meter and smaller sizes, our results indicate that the current mass budget of the swarm at 100~m sizes does not require an outsize parent to explain it.  

\end{abstract}



\section{Introduction} \label{sec:intro}

The Taurid meteor shower is an unusual one. It has a very long duration (approximately 6 months) and contains above-average sized particles, some up to a meter in size \citep{spubormuc17}. The main Taurid meteor showers ---the North and South Taurids--- are dynamically connected to a larger complex of weaker showers \citep{stopor90}, the whole seemingly connected to the unusual comet 2P/Encke \citep{whi67}. These features together with the substantial total mass of the stream ($10^{13}$~kg, \cite{yejen22}), led to it being proposed as the remnant of a giant comet breakup by \cite{clunap84b}.

This hypothesis suggests that a particularly large (hundred km scale) comet was delivered to and broke apart in the inner solar system 10–20 thousand years ago, and that the Taurids and 2P/Encke are the principal remnants. \cite{ashclu93} further proposed that many of the fragments were captured into and dynamically protected by the 7:2 mean motion resonance (MMR) with Jupiter, forming what they called the Taurid Resonant Swarm (TS). It was also proposed that there could be an increase in Taurid meteor activity, particularly at larger ($1$~m possibly up to 100~m sizes) impacting Earth at the times when our planet happens to pass through the TS \citep{ash91, ashclu93, ashclunap94}.

An extended version of this hypothesis, known as "Coherent Catastrophism" posits that the Taurid complex is the dominant source of Earth impactors at tens to hundreds of meter sizes \citep{ashclunap94}. This proposal is not universally accepted. 
\cite{valmorgon95} pointed out the possibility of coincidental orbital similarities between the Taurids and unrelated NEAs. The past orbital history of the Taurid complex was investigated in detail by \cite{egawiebro21} who found that some asteroids on Taurid-like orbits might be dynamically linked, but that spectral observations would likely be needed to resolve whether or not they are from the same parent. In fact, many asteroids previously thought linked to the Taurids on the basis of orbital similarity \citep{ashcluste93, steash96} have subsequently been found to have spectra that differ from each other and/or from 2P/Encke \citep{tubsnomic15, popbirned14}, which argues against a genetic relationship.

However, there is also some evidence in support of the existence of the TS. Increased rates of seismically detected lunar impacts were reported in 1975 during a time of Earth's passage near the TS \citep{obenak87}. Increased meteor activity has also occurred during TS passages \citep{ashizu98,beeharbro04,egabrowie22a}. In particular
a 2015 Taurid outburst occurred, resulting in more than 100 bright Taurid fireballs (dm-sized up to a meter) being observed by the European Meteor Network \citep{spubormuc17}. The resulting data showed that most of the fireballs were indeed strongly associated with the 7:2 MMR with Jupiter.

Observational searches for Taurid swarm members in space have so far been unsuccessful. A 2019 opportunity \citep{clawiebro19} went untapped when protests at Maunakea prevented telescope operations at the Canada-France-Hawaii Telescope (CFHT) during the observing window. Here we report on the results of 2022 observing campaign also attempted at CFHT, in this case where imaging was successfully obtained. Another search for TS members during the 2022 apparition by the Zwicky Transit Facility reported no detections \citep{liyevid25}.

\section{Methods}
\subsection{Observations} \label{sec:observations}
Images were obtained from the CFHT with the MegaCam imager on three nights (29 to 31 October 2022) through the \textit{gri} filter.
Four separate pointings were arranged along on the expected direction of motion of the Taurids, with the hope of doing self follow-up of any Taurids imaged, as they were expected to be too faint to follow-up easily with other telescopes.
For each pointing, three dithered 260 second sidereally-tracked exposures were taken before moving to the next pointing. With MegaCam's 40-second image download time, this resulted in each pointing acquiring three images over the course of 15 minutes, before the telescope proceeded to the next pointing. All four pointings could be completed in an hour with the sequence restarted if conditions allowed, for a total of two hours of observation. Only the first night saw the full 2 hour sequence completed, but substantial numbers of images were taken on subsequent nights which allowed for significant additional coverage.

\begin{figure}
    \centering
    \includegraphics[width=0.85\textwidth]{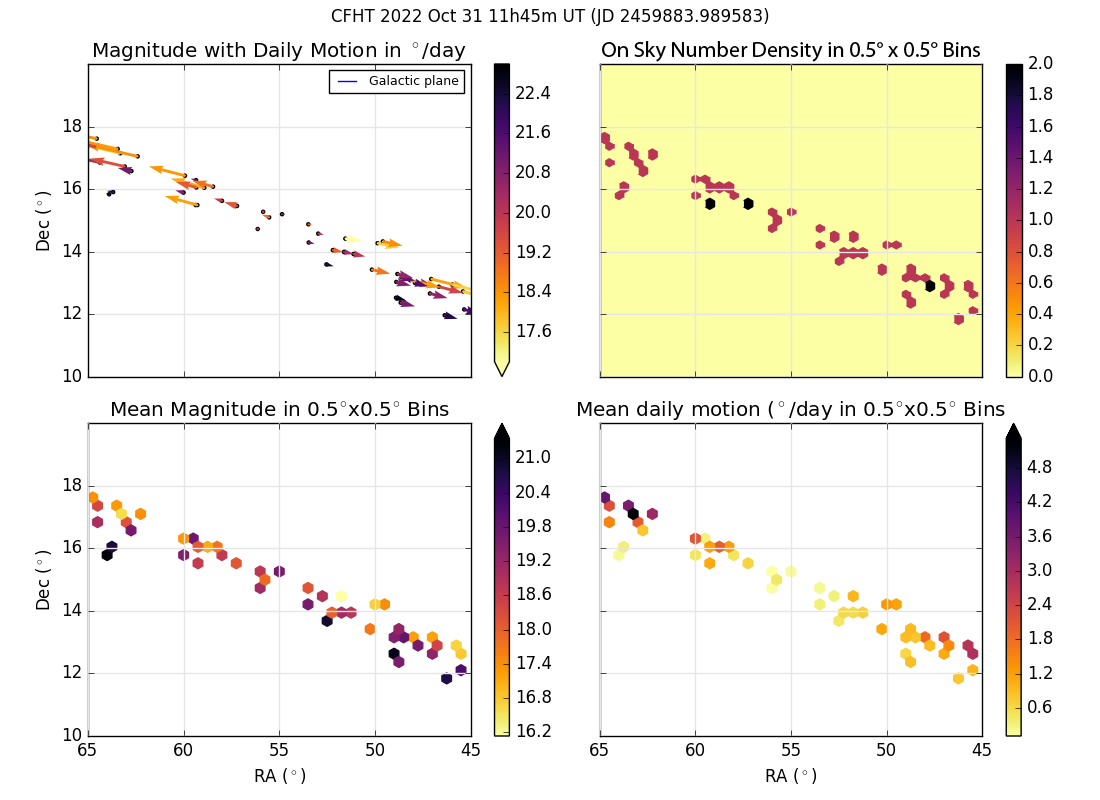}
    \caption{An example of modelled on-sky locations and motions of the Taurid Swarm used to plan observations. Each panel shows a portion of the sky, and the Taurid radiant is located near the center of the plot. Observations were chosen to cover a portion of the sky extending from the center to the lower right. The upper left panel shows the location of simulated Taurids colored by their apparent magnitude along with an arrow indicating their direction of motion. The lower left panel shows the mean apparent magnitude per bin; the upper right panel shows the on-sky number density; and the lower right panel indicates the mean binned on-sky motion. See the text for further details. }
    \label{fig:onsky-model}
\end{figure}

Observations were directed towards the point where the on-sky motion of the Taurids would be minimized, essentially near the radiant of the Taurid meteor shower near a Right Ascension and Declination of ($55\degree$, 15$\degree$). Modelled Taurids were distributed as in \cite{clawiebro19}. Figure~\ref{fig:onsky-model} shows the modelled on-sky motion of the Taurid Swarm used to plan observations. The upper left panel shows the location of simulated Taurids colored by their apparent magnitude, along with an arrow indicating their direction of on-sky motion. Objects on the left hand side of the plot are moving to the left and vice-versa. The rates of motion typically increase the further one goes from the center, which is also illustrated in the lower right panel which shows the mean daily motion in discrete bins. These two panels indicate that observing regions close to the center of the figure should capture objects with the lowest on-sky rates of motion, and that was the strategy we adopted. Observations were chosen to cover a portion of the sky extending from the center to the lower right of this portion of sky. The lower left panel shows the mean apparent magnitude per bin, and the upper right panel shows the magnitude-weighted on-sky density. These are relatively uniform and indicate that object brightness does not vary much across the region in question. Although the Taurids would be brighter in other parts of the sky, our limited angular coverage led us to concentrate on where the Taurids should be most densely concentrated, at some loss of apparent magnitude. 

\subsection{Moving object detection}

The moving object detection pipeline (which is described in \cite{gilwie09,gilwie10}) flagged triplets of sources moving in any direction at on-sky rates of up to 150 arcseconds per hour. Each detection was verified by a human operator. Based on past experience with searches for moving objects at similar rates we expect a detection efficiency of 80\%. The detection code requires the source to be detected in all three frames, and so the 20\% of lost objects typically occur due to sources moving into chip gaps or onto stars.

A total of 8739 individual moving sources were detected in the images, which were ultimately associated with 1433 unique minor planets, both known and unknown. Eight candidate Taurids were flagged for further analysis (discussed further in section~\ref{sec:candidates}).

\begin{figure}
    \centering
    \includegraphics[width=0.85\textwidth]{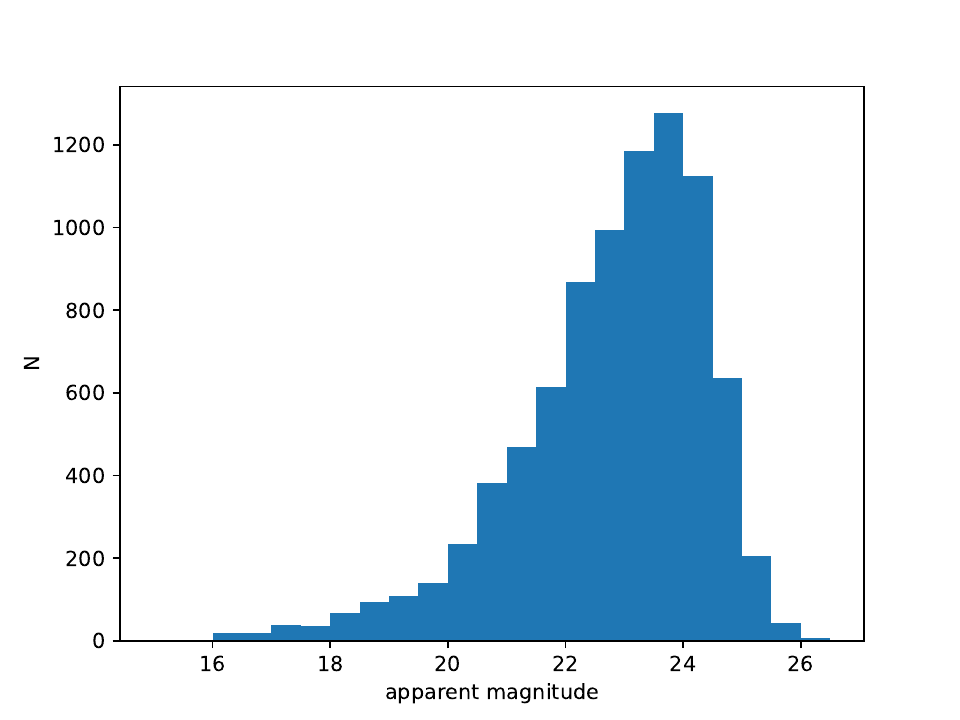}
    \caption{The distribution of apparent magnitudes of detected moving objects.}
    \label{fig:appmag}
\end{figure}

A histogram of the apparent magnitudes of the individual moving object detections is presented in Figure~\ref{fig:appmag}. The overall 50\% detection limiting magnitude is 24.5. From our simulations (see Section~\ref{sec:upperlimit}) we find that any Taurids within our CFHT images would be at distances $\Delta = 0.33 \pm 0.04$ au from Earth at low phase (10-20 deg). Assuming a photometric $G=0.15$, a Taurid with a given absolute magnitude $H$ would have an apparent magnitude simply offset so that $m=H-1.1\pm0.3$. Therefore our limiting apparent magnitude of $m < 24.5$ translates roughly to an absolute magnitude limit of $H < 25.6 \pm 0.3$. Assuming an Encke-like albedo ($0.046 \pm 0.023$, \cite{camfer02}) and the use of the standard HG photometric approach \citep{bowhapdom89}, our observational limiting magnitude corresponds to a limiting diameter of $47^{+29}_{-13}$~meters.

\subsection{Detections of interest} 
Many non-Taurid objects, in particular main-belt asteroids, are present in our images. By design, the observing circumstances allow the majority of main belt objects to be excluded from further analysis based simply on their on-sky rates of motion. To illustrate this, we present in Figure~\ref{fig:meanmotion} the on-sky expected rates of motion for asteroids in the main belt as well as Taurids at the time our images were obtained.

The rates of motion were determined from a numerical integration of hypothetical particles distributed to approximate the main belt and the Taurid stream, and then selected based on the observing geometry applicable to the images taken. Main-belt asteroids were simply distributed on random orbits with semi-major axes $a$ between 2 and 4 au, eccentricities $e$ between 0 and 0.5, and inclinations $i$ between 0 and $60\deg$. This is not intended to provide an accurate description of the main belt but rather to provide a very broad sample for comparison. Taurid on-sky motions were derived from the model described in Section~\ref{sec:observations}.

Figure~\ref{fig:meanmotion} demonstrates that most main belt objects can be quickly filtered out on the basis of their on-sky motion. There is only a small region of overlap where Taurids and main belt objects have similar rates of motion at large negative rates in Declination. This allows us to rapidly reduce our list to those events most likely to be Taurids. We examine more carefully all the candidates which fall below the dashed line on Figure~\ref{fig:meanmotion}. 
Only eight of our $>1000$~moving objects are consistent with Taurids: they are discussed in the next section.

\section{Results}
\begin{figure}
    \centering
    \includegraphics[width=0.85\textwidth]{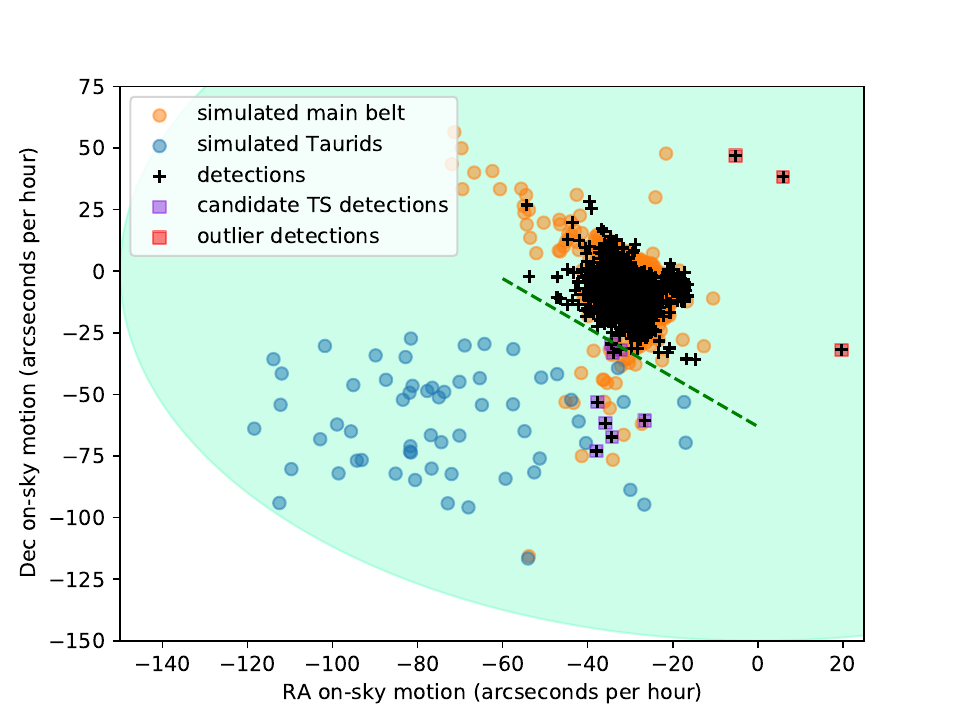}
    \caption{The expected on-sky motion of simulated main belt asteroids and Taurid stream members within the images taken. The on-sky motions of all detected moving objects are superimposed. The rates of on-sky motion that pass our moving-object detection filter (that is, on-sky rates of motion less than 150 arcseconds per hour) are indicated by the green circular area. The moving objects falling below the dashed line are our Taurid candidates (indicated in purple). These eight, which include three near the boundary together with five with larger negative rates of motion in Declination are discussed in Section~\ref{sec:candidates}.}
    \label{fig:meanmotion}
\end{figure}

Eight moving objects were detected with on-sky motions within the region where the Taurid and main belt rates overlap. There are also three outliers that appear on the right-hand side of Figure~\ref{fig:meanmotion}. 

\subsection{Candidates and outliers}
\label{sec:candidates}
The eight moving sources discussed below had rates of motion on the sky consistent with Taurids but careful analysis was able to eliminate seven of them. Details are below. 
\begin{itemize}
\item The three candidates near the boundary are ALA35F ($m_{gri}=22.3$), ALA36W ($m_{gri}= 20.6$) and ALA3WE ($m_{gri}= 21.5)$ . These are asteroids 
2017 GV35, (549797) 2011 SH284 and (566030) 2017 KA33 respectively, which are in the main belt. 
\item Candidate ALA3D3 ($m_{gri}=22.3$) was seen in two triplets of images on a single night (29 Oct 2022) by our survey. It was subsequently linked with Pan-STARRS1 and Pan-STARRS2 data. Now designated 2022 US$_{160}$ it has a main belt orbit, and is not a Taurid.
\item Candidate ALA3e1 ($m_{gri}=22.6$) was seen by our survey only in a single triplet of images on a single night (31 Oct 2022). It is not a Taurid but rather a Hungaria now designated 2022 UV$_{123}$ that was first observed by Pan-STARRS2.
\item Candidate ALA3dh ($m_{gri}=21.7$) was seen in our survey in a single triplet on 31 Oct 2022. It is also a Hungaria, now designated 2022 UT$_{123}$ and provisionally discovered by the Catalina Sky Survey (G96).
\item ALA3MC was observed in two image triplets over two nights (30 and 31 Oct 2022) in our survey. With an apparent magnitude of $m_{gri} = 23.3$ it was not found in a search of Pan-STARRS images. Its Digest2 \citep{keyverpay19} scores are 10 and 20 on each night and its nominal orbit indicates a probable Mars Crosser or a near-Mars crosser. It is not a Taurid.
\item Candidate ALA3gK was seen in only one triplet on 31 Oct 2022. A search for it in Pan-STARRS data was unsuccessful and it was so faint that observations by other stations are unlikely ($m_{gri} = 23.9 \pm 0.5$). 
Its orbit based on the short observational arc is ambiguous, though its rates of motion (see Figure~\ref{fig:meanmotion}) are consistent with it being a Taurid. Though we conclude it is unlikely to be a TS member, if it was a Taurid, using a typical distance to the TS swarm during the observations and assuming an Encke-like albedo, it would be $H = 25.0$ and have a nominal diameter of 60 meters.
\end{itemize}

The few outliers in rates of motion on Figure~\ref{fig:meanmotion} are less common objects but are not Taurids. 
\begin{itemize}
\item Candidate ALA3h7 ($m_{gri}=21$) was seen in two triplets on two nights (30 and 31 Oct 2022) and is a Mars crosser now designated 2022 WL$_1$.
\item Candidate ALA3Jy ($m_{gri}=24.5$) was seen in two triplets on a single night (30 Oct 2022).  Too faint to find in PanSTARRS archival data, its Digest2 score \citep{keyverpay19} of 78 suggests it might be a Near-Earth Object (NEO). Its rates of motion are not compatible with it being a Taurid.
\item Candidate ALA3To ($m_{gri}=23.1$) was seen in three triplets on three separate nights (29, 30 and 31 Oct 2022). It is not a Taurid but rather a Mars crosser now designated 2022 UT$_{160}$. 
\end{itemize}

\begin{figure}
    \centering
    \includegraphics[width=0.85\textwidth]{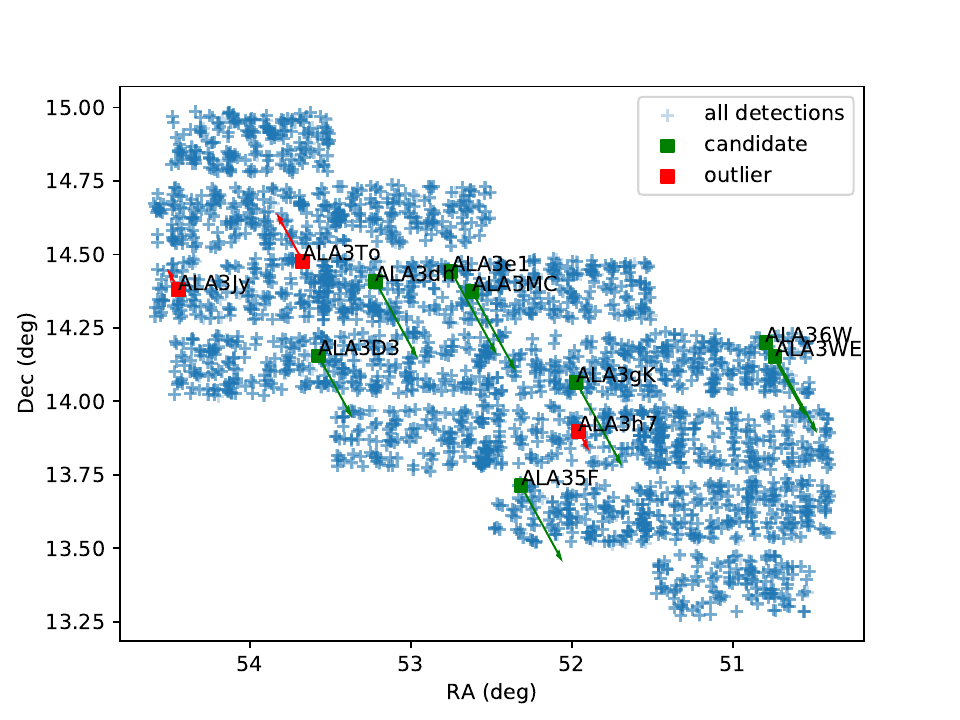}
    \caption{The locations of all object detections on the sky. Our four slightly overlapping CFHT MegaCam fields are aligned along the expected direction of motion of Taurids. The eight candidates TS objects discussed in section~\ref{sec:candidates} superimposed, as are the three additional outliers in on-sky motion. Arrows are directed along the directions of motion. The gaps between detections correspond to the gaps between the chips of the  MegaCam detector.}
    \label{fig:RADec}
\end{figure}

\subsection{Upper limit on population of Taurid Swarm} \label{sec:upperlimit}
\begin{figure}
    \centering
    \includegraphics[width=0.85\textwidth]{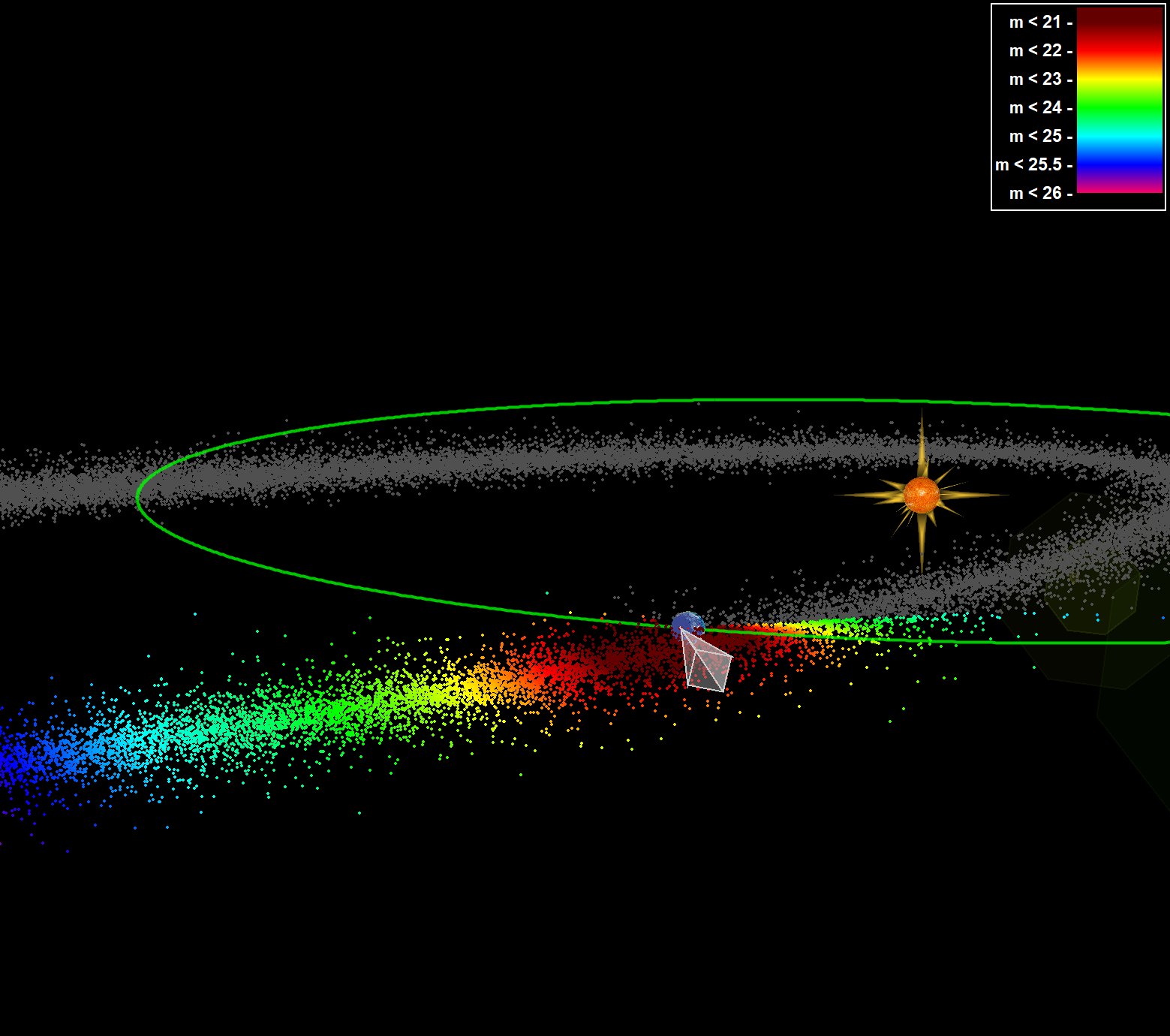}
    \caption{The geometry of the Earth relative to the Taurid stream at one instant during the survey. The white frustum indicate schematically the volume of space sampled by the observations. The color of the Taurid particles indicates their apparent magnitude as seen from Earth.}
    \label{fig:geometry}
\end{figure}

This survey was deliberately timed to occur when the Earth was passing close to the middle of the Taurid stream. Figure ~\ref{fig:geometry} illustrates the relative geometry during the survey. At this time, the relatively small area of our survey is compensated for by the relative concentration of Taurids on the sky as seen from the Earth's vantage point. However we still only sample a small fraction of the TS cloud. But how much?
The number of TS members detectable by our survey is a complicated function of their orbital distribution, sizes and the observing circumstances. Nonetheless we can set some limits on the population of the Taurid swarm by considering some simple limiting cases. Here we will examine two hypothetical TS populations, populations designed to bracket the real one to the extent possible.

The first case (which we will call the "broad" scenario) is where the TS population is assumed to be distributed over entire orbital phase space that is in 7:2 resonance with Jupiter. This spreads the TS mass over the largest volume. The scenario is modelled via a set of hypothetical TS particles generated in the work of \cite{clawiebro19} without any restriction to the "core" of the swarm.  The second case ("narrow" scenario) we will consider is where the TS population is confined in the portion of the 7:2 resonant phase space which is known to be populated by fireball producing material. This scenario is modelled via a set of hypothetical TS particles distributed within the range of orbital elements observed during the 2015 outburst, when 144 Taurid fireballs were observed by the European Fireball Network \citep{spubormuc17}. The narrow TS is populated only by particles within $\pm 35 \degree$ of the swarm center, representing the predicted extent of the TS (\cite{ashclu93}; see later in this section for more on the predicted extent of the TS in mean anomaly). By simulating the positions and motions of these two hypothetical TS populations on the dates of observation we can determine what fraction of the TS swarm would have been observed from CFHT in each case. 

In the "broad" case, we find that only 1 in 8000 TS members would have been within our observing volume, while in the "narrow" case, that number increases by an order of magnitude, because our telescopic search area is concentrated on this notional center of the TS. Given that the Poisson 95\% confidence range for zero detections is [0,3.69] \citep{meehahesc17}, we can therefore set an upper limit of fewer than $3 \times 10^3$ to $3 \times 10^4$ TS members at 50 meter diameters, these numbers corresponding to the broad and narrow TS models respectively. If all the TS mass is at these sizes, then this corresponds to an equivalent progenitor body of only 50 m~$\times (N)^{1/3}= 700 - 1500$~m diameter, far short of the 50-100~km size body proposed in the past \citep{clunap84b}. 

How much TS mass could be hiding at larger sizes? From recent estimates of the completeness of near-Earth object (NEO) catalogs by \cite{gramaimas23}, it is thought that the catalog is 88\% complete at km-sizes and 38\% at 140~m. Thus it is unlikely (though not impossible) that a km-class TS member remains undetected, while undetected 100-m class members remain a distinct possibility. If for argument's sake we assume that all the mass in the TS is in 140 m class objects, our non-detection here only requires of the total size of the progenitor body a size of 0.14~km~$\times (N)^{1/3}= 2-4.5$~km diameter.

There are other possibilities which could result in our calculations above underestimating the mass of the TS. The first is that the mass could be highly localized within the swarm, and our observations examined a region with little or no mass. However, fireball outbursts have been reported for each of the November TS swarm returns predicted since 1988, namely 1988, 1998, 2005 and 2015 \citep{egabrowie22a} as well as 2018 and 2022 \citep{spubor23}. During these returns the Earth passed within 5, -13, 11 -7, -48, and 17 degrees of mean anomaly respectively from the predicted TS center. A positive value indicates that the center of the swarm is past the Earth at the time in question and vice-versa for negative values\footnote{\url{https://www.cantab.net/users/davidasher/taurid/swarmyears.html} retrieved 2024 Oct 13. See \cite{ashclu93} for more details.}. These outburst observations suggest that the fireballs are distributed more or less evenly across at least a 30 degree stretch of mean anomaly, with 2018 being the only exception when the mean anomaly offset was as high as -48 degrees. The observations reported here were taken when the Earth was at 17 degrees of mean anomaly from the TS center. During this 2022 apparition over 150 Taurid fireballs ---the "vast majority" associated with the 7:2 resonance --- were recorded by the European Fireball Network \citep{spubor23}. Thus it seems clear that Earth was within the TS during the time our observations were collected. However we were also looking towards the outer edge of the swarm (of necessity, the swarm center was in the daytime sky, see Figure~\ref{fig:geometry}) which much reduces the volume of the TS we sampled. 

Another possibility is that the Taurid Swarm contained larger objects in the past, and that these have decayed through the loss of volatiles or other processes into smaller ones.  Comets are known to split:  \cite{boe04} report that  10 of the 160 short-period comets known at the time had been observed to fragment at some level. The relatively low perihelion distance $q\approx 0.3$~au of the Taurid stream would also promote loss of volatiles even without macroscopic fragmentation. The observed Jupiter-family comet distribution in general shows fewer members at sizes below $\sim 1$ km than would be expected from a simple power law extrapolation from larger sizes  \cite[and refs therein]{snofitlow11}. This could be due to observational biases but others have argued it is not entirely due to such effects \citep{meehaimar04}. It is possible therefore that the Taurid swarm contained more mass in the past, though observations do not show a need for much hidden mass at the current time.

Thus we conclude that while fireball observations do confirm the existence of a Taurid Swarm at small sizes (meter-sized and smaller), negative observations at telescopic sizes suggest that there is no need for the break-up of a particularly large parent body, but that the TS total mass is largely constrained to be no more than that of a more-or-less typical comet.

\section{Conclusions}

We conducted a deep narrow survey of the Taurid Swarm during its 2022 apparition. No definite Taurid Swarm members were detected. One possible TS candidate was seen but its orbit was ambiguous from the observational arc obtained, and a search for follow-up observations reported by Pan-STARRS and other stations was unsuccessful. Our results are consistent with no Taurids detected down to our apparent limiting magnitude of 24.5, which sets our overall upper limit on the population of the TS swarm at fewer than $3 \times 10^3- 3 \times 10^4$ objects (95\% confidence) down to diameters of $47^{+29}_{-13}$ meters assuming an Encke-like albedo. Our results suggest that while fireball observations confirm the Taurid Swarm's existence, the mass in the swarm at 50-100 m sizes is limited to total values below that of a typical comet or asteroid. Though a larger body such has been proposed in the past could be the parent of the TS, the current mass budget of the swarm does not require an outsize parent to explain it.  


\section{Acknowledgments}
We thank Mark Boslough and Peter Brown for helpful discussions that
much improved this work. The authors acknowledge the sacred nature of
Maunakea and appreciate the opportunity to observe from the
mountain. The Canada-France-Hawaii Telescope (CFHT) is operated by the
National Research Council (NRC) of Canada, the Institut National des
Sciences de l’Univers of the Centre National de la Recherche
Scientifique (CNRS) of France, and the University of Hawaii. 
Pan-STARRS is supported by the National Aeronautics and Space
Administration under Grants 80NSSC18K0971 and 80NSSC21K1572 issued
through the SSO Near-Earth Object Observations Program. This research
has made use of data and/or services provided by the International
Astronomical Union's Minor Planet Center. This
study was supported in part by the NASA Meteoroid Environment Office
under Cooperative Agreement No. 80NSSC24M0060 and by the Natural
Sciences and Engineering Research Council of Canada (NSERC) Discovery
Grant program (grants No. RGPIN-2018-05659 and RGPIN-2024-05200).

%

\vspace{5mm}
\facilities{CFHT (MegaCam), Pan-STARRS}





\bibliography{taurids-2022}{}
\bibliographystyle{aasjournal}



\end{document}